\documentclass{article}

    \PassOptionsToPackage{square,sort,comma,numbers}{natbib}
    \usepackage[final]{neurips_2024}

\usepackage[utf8]{inputenc} % allow utf-8 input
\usepackage[T1]{fontenc}    % use 8-bit T1 fonts
\usepackage{hyperref}       % hyperlinks
\usepackage{url}            % simple URL typesetting
\usepackage{booktabs}       % professional-quality tables
\usepackage{amsfonts}       % blackboard math symbols
\usepackage{nicefrac}       % compact symbols for 1/2, etc.
\usepackage{microtype}      % microtypography
\usepackage{xcolor}         % colors
\usepackage{graphicx}
\usepackage{listings}
\usepackage{float}
\usepackage{enumitem}

\title{Vision Language Models Are Few-Shot Audio Spectrogram Classifiers}

\author{
\quad  Satvik Dixit \quad Laurie M.  Heller \quad Chris Donahue\\
Carnegie Mellon University \\
\texttt{\{satvikdixit, laurieheller, chrisdonahue\}@cmu.edu}
}

\begin{document}

\maketitle

% \footnotetext{This work was generously supported by Sony via the Faculty Innovation Award program.}

\begin{abstract}
We demonstrate that vision language models (VLMs) are capable of recognizing the content in audio recordings when given corresponding spectrogram images. 
Specifically, we instruct VLMs to perform audio classification tasks in a few-shot setting by prompting them to classify a spectrogram image given example spectrogram images of each class. By carefully designing the spectrogram image representation and selecting good few-shot examples, we show that GPT-4o can achieve $59.00$\% cross-validated accuracy on the ESC-10 environmental sound classification dataset. Moreover, we demonstrate that VLMs currently outperform the only available commercial audio language model with audio understanding capabilities (Gemini-1.5) on the equivalent audio classification task ($59.00$\%~vs.~$49.62$\%), 
and even perform slightly better than  human experts on visual spectrogram classification ($73.75$\%~vs.~$72.50$\% on first fold). 
We envision two potential use cases for these findings: 
(1)~combining the spectrogram and language understanding capabilities of VLMs for audio caption augmentation, and
(2)~posing visual spectrogram classification as a challenge task for VLMs.
\end{abstract}

\section{Introduction}

% VLMs are able to do really well on a lot of tasks
Vision-language models (VLMs) have emerged as a powerful paradigm for multimodal artificial intelligence, capable of jointly processing and reasoning over visual and textual information \cite{Du2022ASO, bordes2024introductionvisionlanguagemodeling}. By integrating computer vision and natural language processing capabilities, VLMs have demonstrated remarkable capabilities across a wide range of tasks, from image captioning and visual question answering to object detection and scene understanding \cite{zhang2024vision, jiang2024effectivenessassessmentrecentlarge}. The success of VLMs in these domains suggests that their generalized visual reasoning abilities could be leveraged for tasks in other modalities, supposing those modalities can be reasonably represented as images. 

% Some motivation behind the idea
Here we investigate whether VLMs might be able to perform audio classification when presented with audio in the form of \emph{spectrograms}. 
Spectrograms, which visually represent the frequency content of audio signals over time, have long been used as input for audio classification tasks \cite{piczak2015esc, choi2016convolutionalrecurrentneuralnetworks, hershey2017cnnarchitectureslargescaleaudio, Pons2017EndtoendLF,gong2021astaudiospectrogramtransformer}. 
These time-frequency representations capture essential acoustic features that are often more informative than raw waveforms for many audio analysis tasks. 
Recent commercial VLMs, such as GPT-4o \cite{gpt4}, Claude-3.5 Sonnet \cite{anthropic2024claude3} and Gemini-1.5 \cite{gemini}, have demonstrated impressive zero-shot and few-shot capabilities across various visual tasks \cite{yue2024mmmumassivemultidisciplinemultimodal}. These models are trained on image-text data and accordingly have seen spectrograms and associated text during pre-training. However, their potential for processing spectrograms and classifying audio content remains unexplored to the best of our knowledge. 

We propose a novel task called \emph{visual spectrogram classification} (VSC) which tasks models with recognizing the content in audio recordings from their equivalent visual spectrograms. We benchmark VLMs and attempt to measure human expert performance on this task. Our approach involves carefully designing the spectrogram image representation and the few-shot prompting strategy to enable the VLM to reason about audio content from visual patterns. 

We conduct comprehensive experiments to evaluate the zero-shot and few-shot performance of various commercial VLMs on this task. We conduct ablation studies to optimize spectrogram hyperparameters, exploring variations in amplitude scale, frequency axis, colormap, and spectrogram style transformations. We also experiment with different numbers and types of examples for few-shot learning. Finally, we compare the few-shot performance of VLMs with the performance of human experts on VSC and commercial audio language models on audio classification (AC).

% We also compare the cross-validated VSC performance of GPT-4o to the cross-validated audio classification (AC) performance of state-of-the-art commercial and opensource audio language models (Gemini-1.5 and Pengi) on the ESC-10 dataset.

% Why: Audio Caption Augmentation
Demonstrating that VLMs can comprehend sounds through spectrograms opens up a range of potential applications. One of our aims, for instance, is to improve the quality of captions in audio captioning datasets which often contain vague or inaccurate descriptions \cite{Drossos2019ClothoAA}. While large language models (LLMs) are capable of generating captions, they are prone to hallucination. Incorporating audio information would help the multimodal LLM to ground the captions but it is not immediately clear how to do so. Current audio language models (ALMs) like Pengi \cite{deshmukh2024pengiaudiolanguagemodel}, SALMONN \cite{tang2024salmonngenerichearingabilities} and GAMA \cite{ghosh2024gamalargeaudiolanguagemodel} allow audio as input but lag behind commercial VLMs in terms of language understanding abilities, as evidenced by the size of their underlying LLMs (Pengi uses GPT-2 as the language model component with 124M parameters compared to the hundreds of billions now common in state-of-the-art models \cite{dubey2024llama}). If we could provide both the vague caption and the audio information to the VLM as a spectrogram, we may be able to get captions that are both grounded in audio as well as sufficiently descriptive.

% Larger goal
% Furthermore, our research contributes to the growing body of work on cross-modal learning and the adaptability of VLMs to novel tasks. 
Our research contributes to cross-modal learning and VLM adaptability to novel tasks.
By demonstrating the effectiveness of VLMs in processing spectrograms for audio classification, we highlight the potential for these models to bridge the gap between visual and auditory domains. VSC serves as a new benchmark task for VLMs as a test of their audio spectrogram understanding capabilities.

% Contributions
The primary contributions of this work are:
\begin{itemize}
\item Proposing a novel task, Visual Spectrogram Classification (VSC), which demonstrates VLMs' capability to classify audio content using spectrogram images.
\item Conducting ablation studies to find optimal spectrogram hyperparameters for VSC. 
\item Comparing the performance of latest models from the GPT-4, Claude and Gemini series on the VSC task in zero-shot and few-shot settings.
\item Comparing VLM performance with human experts on the VSC task and commercial audio language models on the audio classification task.
\end{itemize}

\section{Tasks \& Methods}

\subsection{Task definition for visual spectrogram classification (VSC)}
Visual spectrogram classification (VSC) is a novel task that involves classifying audio content based on the spectrogram representations. In zero-shot settings, the model analyzes the spectrogram and selects the most likely audio class. For few-shot settings, the model is provided with example spectrograms for each sound class to guide the classification process.   

% We introduce Visual Spectrogram Classification (VSC), a task where models classify audio content using spectrogram representations. In zero-shot settings, models analyze spectrograms to select the most probable audio class. Few-shot settings provide example spectrograms for each sound class to guide classification.

\subsection{Default spectrogram extraction hyperparameters}
% To ensure consistency, we establish a set of default hyperparameters for spectrogram extraction. Audio files are resampled to 22,050 Hz and Short-time Fourier transform (STFT) is computed using a window size of 2,048 samples and a hop length of 512 samples. Both the frequency axis and the amplitude scale are set to logarithmic as is common in audio research. The `viridis' colormap is chosen as default. To improve clarity, we add axis labels and remove the colormap scale from the spectrogram images.

We establish default parameters for spectrogram extraction to ensure consistency. Audio files are resampled to 22,050 Hz, with Short-time Fourier transform (STFT) computed using a 2,048-sample window size and 512-sample hop length. Both frequency and amplitude scales are logarithmic (as is common in audio research), using the 'viridis' colormap. To improve clarity, we added axis labels and removed the colormap scale from the spectrogram images.

\subsection{Prompting VLMs to perform VSC}
 % For zero-shot experiments, we use the spectrogram and a text prompt as input to the model. The text prompt lists all the classes and asks the model to choose the most likely class based on the spectrogram. For few-shot experiments, we utilize in-context learning \cite{dong2024surveyincontextlearning} by also providing the model one example spectrogram of each class. 
 
 For zero-shot experiments, we input the spectrogram and a text prompt listing all classes, instructing the model to select the most likely class. Few-shot experiments employ in-context learning \cite{dong2024surveyincontextlearning}, providing one example spectrogram per class alongside the test spectrogram.

\section{Experiments \& Results}

\subsection{Dataset and Models}

% The experiments are conducted using the ESC-10 dataset \cite{piczak2015dataset} which contains 5s environmental sound recordings. It is a  subset of the ESC-50 dataset and contains 400 audio clips across 10 classes - chainsaw, clock tick, crackling fire, crying baby, dog, helicopter, rain, rooster, sea waves and sneezing. This dataset is divided into five folds with each fold containing 80 audio clips (8 clips per class). The spectrograms for one example from each of these 10 classes is shown in \ref{fig:example_spectrograms}

\begin{figure}[h!]
    \centering
    \includegraphics[width=1\linewidth]{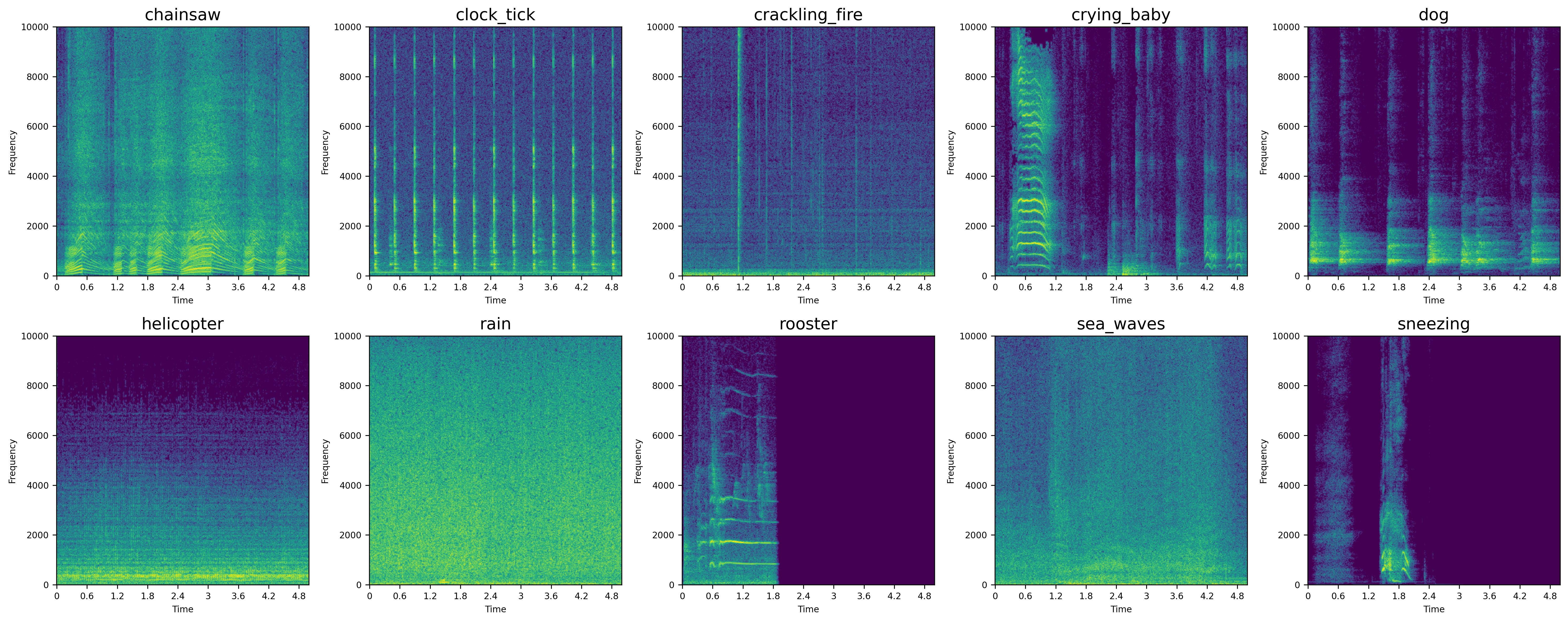}
    \caption{Example audio spectrograms for each class in the ESC-10 dataset}
    \label{fig:example_spectrograms}
\end{figure}

\textbf{Dataset}: We utilize the ESC-10 dataset, a subset of ESC-50 \cite{piczak2015esc}, comprising 400 5-second environmental sound recordings across 10 classes. The dataset is divided into five folds, each containing 80 audio clips (8 per class). Figure \ref{fig:example_spectrograms} illustrates example spectrograms for each class.

% \subsection{Models}

% We use two GPT-4 variants: GPT-4o (most powerful) and GPT-4o-mini (lightweight for faster inference). From Anthropic, we include Claude-3.5 Sonnet (latest) and Claude-3 Opus (computation-intensive). Google’s Gemini series models we include are Gemini-1.5 Pro (most powerful) and Gemini-1.5 Flash (optimized for speed). 
% All models were accessed via API endpoints using the latest versions available as of September 2024, with default API parameters unless noted.
\textbf{Models}: Our experiments employ six state-of-the-art VLMs:
\begin{itemize}[itemsep=0pt]
\item GPT-4: GPT-4o (most powerful) and GPT-4o-mini (lightweight)
\item Claude: Claude-3.5 Sonnet (latest) and Claude-3 Opus (computation-intensive)
\item Gemini: Gemini-1.5 Pro (most powerful) and Gemini-1.5 Flash (lightweight)
\end{itemize}
All models were accessed via API endpoints using the latest versions as of September 2024 %, with default parameters unless specified.

\begin{figure}
    \centering
    \includegraphics[width=1\linewidth]{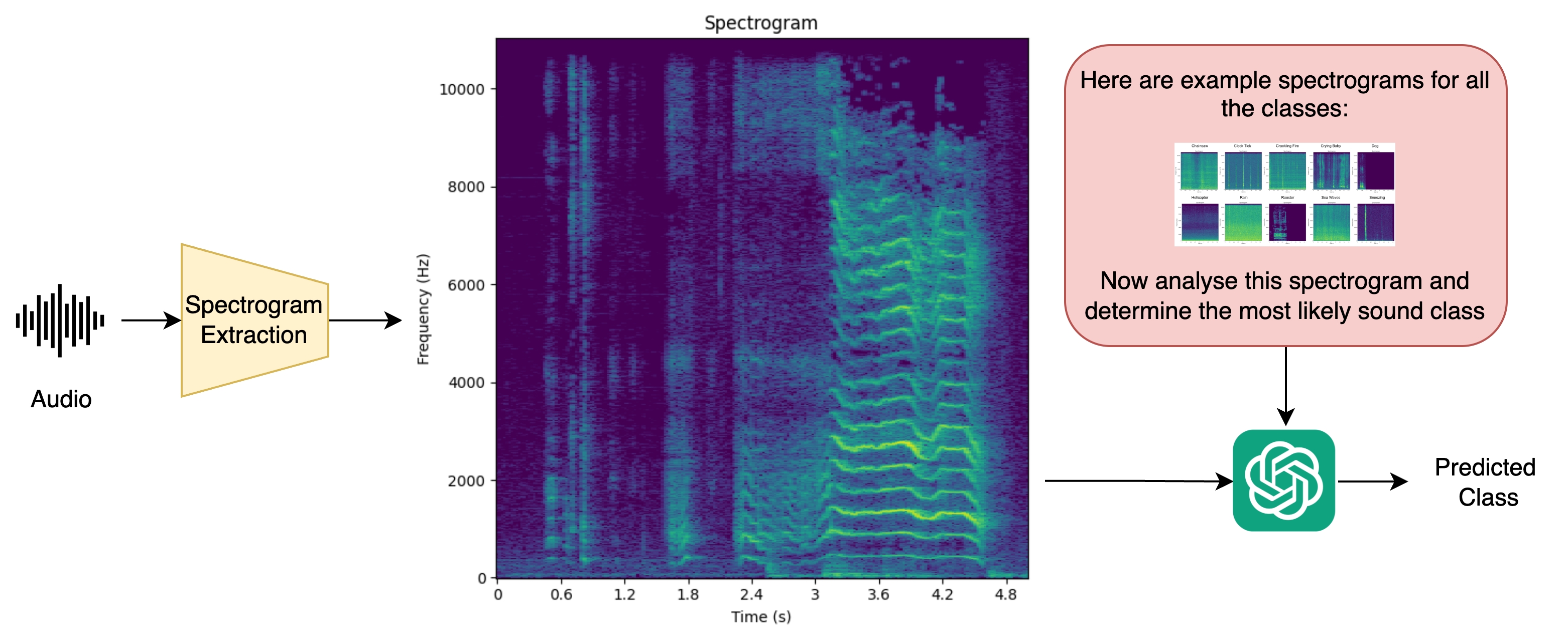}
    \caption{Experimental setup of the visual spectrogram classification task in the few-shot setting}
    \label{fig:fig_1}
\end{figure}

\subsection{Comparing zero-shot performance of VLMs on the visual spectrogram classification task}

We evaluated the zero-shot VSC performance of all the VLMs on the first fold of ESC-10 using the default spectrogram hyperparameters. The first column in Table \ref{table:full} shows the zero-shot VSC performance. GPT-4o outperformed other state-of-the-art models in the zero-shot setting. 
% We evaluated zero-shot VSC performance of all VLMs on the first fold of ESC-10 using default spectrogram hyperparameters. Table \ref{table:full} (first column) shows GPT-4o outperforming other state-of-the-art models in this setting.

\begin{table}[h]
\caption{Zero-shot and few-shot performance of VLMs (on VSC) and ALMs (on AC) on the first fold of the ESC-10 dataset. Zero-shot performance with tuned parameters (see appendix) is also shown. * indicates that the hyperparameters were tuned on GPT-4o, so it has an advantage over other models.}
\label{table:full}
\centering
\begin{tabular}{lp{3.8cm}p{2.5cm}p{2.5cm}p{1.8cm}}  % Adjust the width as needed
\toprule
Modality & Model & Zero-shot (default parameters) (\%) & Zero-shot (tuned parameters) (\%) & Few-shot (\%)\\
\midrule
Vision & GPT-4o & 27.50 & 35.00* & 70.00*\\
& GPT-4o mini & 21.50 & 22.50 & 38.75 \\
& Claude-3.5 Sonnet & 21.50 & 22.50 & 56.25 \\
& Claude-3 Opus & 11.25 & 12.50 & 15.00 \\
% & GPT-4V & - & - & 38.75 \\
& Gemini-1.5 Flash & 16.25 & 17.50 & 10.81 \\
& Gemini-1.5 Pro & 18.75 & 18.42 & 20.31 \\
\midrule
Audio & Gemini-1.5 Flash (Audio) & 27.50 & 27.50 & 40.26 \\
& Gemini-1.5 Pro (Audio) & 38.36 & 38.36 & 44.78 \\
\bottomrule
\end{tabular}
\end{table}

\subsection{Few-shot performance of VLMs on the visual spectrogram classification task}

% In our investigation of GPT-4o's few-shot learning capabilities, we conducted visual spectrogram classification experiments on the first fold of the ESC-10 dataset. Our initial experiment involved randomly selecting one example for each class. Subsequent experiments focused on optimizing the quality of examples provided to the model. To achieve this, we combined the remaining four folds of the dataset and employed K-means clustering to group the mel spectrograms for each class into three clusters. We then selected spectrograms closest to the centroids of these clusters as examples for 10-shot, 20-shot, and 30-shot learning tasks. In order to get better examples, we also tried clustering on amplitude spectrograms and handpicking examples for each class. 

We investigated GPT-4o's few-shot learning capabilities on the first fold of ESC-10. Initial experiments used randomly selected examples for each class. We then optimized example quality by applying K-means clustering to mel spectrograms from the remaining folds, selecting spectrograms closest to cluster centroids for 10-shot, 20-shot, and 30-shot tasks. We also explored clustering on amplitude spectrograms and manual example selection.

% Table \ref{table:full} shows the few-shot performance of all the models in the third column with randomly chosen examples for each class. All models demonstrated significant improvements in accuracy when provided with examples. For instance, 10-shot accuracy of GPT-4o increased to 70\% from 35\% in the zero-shot task. Table \ref{table:example_selection} shows the impact of different number of examples and example selection methods on GPT-4o VSC performance. For the 10-shot scenario, choosing better examples by picking examples close to cluster centroids or handpicking leads to improvements in performance. The optimal performance was achieved with 2 examples per class using K-means clustering on Mel spectrograms (76.25\%). Interestingly, increasing to 3 examples per class led to a slight decrease in accuracy to 70\%.

Table \ref{table:full} (third column) shows few-shot performance with randomly chosen examples. All models demonstrated significant accuracy improvements with examples provided. GPT-4o's 10-shot accuracy increased to 70\% from 35\% in zero-shot.
Table \ref{table:example_selection} illustrates the impact of example selection methods on GPT-4o VSC performance. Cluster-based and handpicked examples improved 10-shot performance. Optimal performance (76.25\%) was achieved with 2 examples per class using K-means clustering on Mel spectrograms. Interestingly, 3 examples per class slightly decreased accuracy. %to 70\%

\begin{table}[h]
\centering
\begin{minipage}{0.6\textwidth}
\centering
\caption{Few-shot VSC performance of GPT-4o on ESC-10 (first fold) with different example selection methods}
\label{table:example_selection}
\begin{tabular}{cccc}
\toprule
\# per class & Selection method & Feature & Accuracy (\%) \\
\midrule
1 & Random & Mel & 70.00 \\
1 & Hand-picked & Mel & 75.00 \\
1 & K-means (k=3) & Mel & 73.75 \\
1 & K-means (k=3) & Amp & 60.00 \\
% 1 & K-means (k=1) & Mel & 55.00 \\
2 & K-means (k=3) & Mel & 76.25 \\
3 & K-means (k=3) & Mel & 70.00 \\
\bottomrule
\end{tabular}
\end{minipage}%
\hfill 
% \hspace{0.05\textwidth} % Adjusts the gap between tables
\begin{minipage}{0.30\textwidth}
\centering
\caption{Few-shot VSC performance of human experts on the ESC-10 dataset (first fold)}
\label{table:human_accuracy}
\begin{tabular}{lc}
\toprule
Expert & Accuracy (\%) \\
\midrule
Expert 1 & 67.5 \\
Expert 2 & 57.5 \\
Expert 3 & 71.25 \\
\midrule
Ensembled & 72.5 \\
\bottomrule
\end{tabular}
\end{minipage}
\end{table}

% \begin{table}[h]
% \caption{Few-shot VSC performance of GPT-4o on the ESC-10 dataset with different number of examples and example selection methods}
% \label{table:example_selection}
% \centering
% \begin{tabular}{cccc}
% \toprule
% \# per class & Selection method & Feature & Accuracy (\%) \\
% \midrule
% 1 & Random & Mel & 70.00 \\
% 1 & Hand-picked & Mel & 75.00 \\
% 1 & K-means (k=3) & Mel & 73.75 \\
% 1 & K-means (k=3) & Amp & 60.00 \\
% % 1 & K-means (k=1) & Mel & 55.00 \\
% 2 & K-means (k=3) & Mel & 76.25 \\
% 3 & K-means (k=3) & Mel & 70.00 \\
% \bottomrule
% \end{tabular}
% \end{table}

\subsection{Human expert evaluations}

% To understand how accurate humans are at this task, we conduct human expert assessments. The experts are graduate students and professors from academic labs who work in audio or speech research. Three such experts are asked to classify all 80 spectrograms from the first fold of the ESC-10 dataset. The experts are also provided one example for each class - the same examples as the VLM in the 10-shot mel spectrogram clustering scenario to ensure consistency. Their individual predictions are then ensembled to determine the best-case accuracy. 

To benchmark human performance on the VSC task, we conducted evaluations with three experts (professors and graduate students from audio or speech research labs). Each expert classified 80 spectrograms from the first fold of ESC-10, using the same 10-shot mel spectrogram examples provided to the VLM 10-shot mel spectrogram clustering scenario for consistency.

\begin{figure}
    \centering
    \includegraphics[width=0.95\linewidth]{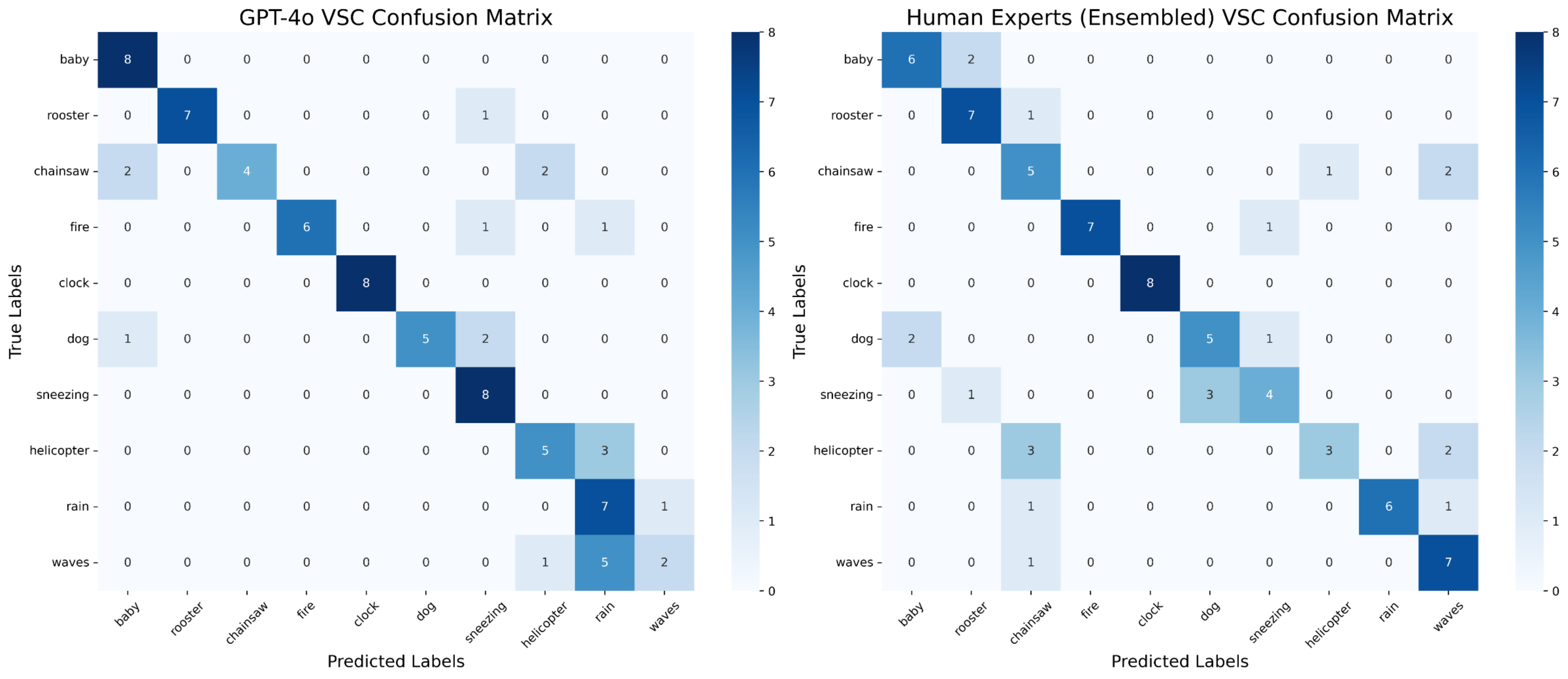}
    \caption{Confusion matrices for GPT-4o (left) and the ensembled human expert predictions (right) }
    \label{fig:fig_cm}
\end{figure}

The accuracies of the experts on the 10-shot VSC task on the first fold of ESC-10 are shown in table \ref{table:human_accuracy}. The mean inter-annotator agreement (Cohen's Kappa Score) is 0.53. Notably, GPT-4o in the same scenario achieved a slightly higher accuracy (73.75\%) than the ensembled human expert performance (72.5\%), demonstrating the model's potential to match and even exceed human expert level performance on this task. The confusion matrices for GPT-4o and ensembled human expert predictions, shown in Figure \ref{fig:fig_cm}, reveal that both approaches struggled more with distinguishing similar spectrograms such as `dog' and `sneezing' or `helicopter', `rain' and `sea waves'.

% \begin{table}[h]
% \caption{Few-shot VSC performance of human experts on the first fold of the ESC-10 dataset.}
% \label{table:human_accuracy}
% \centering
% \begin{tabular}{lc}
% \toprule
% Expert & Accuracy (\%) \\
% \midrule
% Expert 1 & 67.5 \\
% Expert 2 & 57.5 \\
% Expert 3 & 71.25 \\
% \midrule
% Ensembled & 72.5 \\
% \bottomrule
% \end{tabular}
% \end{table}

\subsection{Cross-validated task performance}
To assess the robustness of our approach, we conduct cross-validated VSC performance of GPT-4o and AC performance evaluations of Gemini-1.5 pro on the full ESC-10 dataset. To evaluate GPT-4o on a more challenging task, we use a custom subset of the ESC-50 dataset. ESC-50 is a more comprehensive dataset with 2000 audio files across 50 classes. We create a subset of 100 randomly selected audio files (2 audio files per class) from the first of the five folds for VSC evaluation.
% and evaluate the VSC performance of GPT-4o on it.

The accuracies of the GPT-4o and Gemini-1.5 pro models on the ESC-10 dataset are 59.00\% and 49.62\%  respectively. This highlights the competitive performance of GPT-4o on VSC, even when compared to commercial audio language models. The 50-shot VSC accuracy of GPT-4o on the customized subset of the ESC-50 dataset is 14\% which shows its poor generalization capability when the number of classes increases.

% \begin{table}[h]
% \caption{The cross-validated VSC performance of GPT-4o and AC performance of Gemini and Pengi on the full ESC-10 dataset}
% \label{table:SOTA_models}
% \centering
% \begin{tabular}{lc}
% \toprule
% Model & Accuracy (\%) \\
% \midrule
% GPT-4o (VSC) & - \\
% Gemini (AC) & - \\
% Pengi (AC) & - \\
% \bottomrule
% \end{tabular}
% \end{table}

\section{Limitations}
% While our study demonstrates the potential of VLMs for few-shot audio classification using spectrograms, it is important to acknowledge two significant limitations.
Firstly, current VLMs do not yet match the performance of traditional audio classification methods. This performance gap is likely because VLMs are primarily trained on natural images and text, not spectrograms. Moreover, this task is difficult even for human experts.
   % \chris{I think the fact that this task is difficult even for human experts may be more pertinent}

Secondly, our experiments reveal a significant drop in classification accuracy as the number of audio classes increases (59\% accuracy on the ESC-10 dataset (10 classes) to 14\% accuracy on the ESC-50 dataset (50 classes)). This decline suggests that VLMs struggle to differentiate between a large number of audio classes; the visual patterns in spectrograms may become less distinctive or more confusing to the model as the number of classes or the similarities between them grows.

It is important to note, however, that as VLMs continue to evolve and become larger in size and potentially encounter more spectrograms in their training data, we anticipate improvements in their visual spectrogram classification capabilities.

\section{Conclusion}

In this study, we propose the visual spectrogram classification (VSC) task and evaluate vision language models such as GPT-4o in zero-shot and few-shot settings. Our experiments demonstrate that VLMs can interpret the ability to classify audio spectrograms, with few-shot learning greatly enhancing performance. We show that GPT-4o performs better than commercial multimodal models with audio understanding, and can even outperform human experts on this task. As VLMs continue to evolve, we envision expanded applications in audio classification and understanding.

\newpage

\section{Acknowledgements}
This work was generously supported by Sony via the Sony Research Award Program (RAP).

\bibliographystyle{plain}
\bibliography{references}

\begin{thebibliography}{10}

\bibitem{gpt4}
Josh Achiam, Steven Adler, Sandhini Agarwal, Lama Ahmad, Ilge Akkaya, Florencia~Leoni Aleman, Diogo Almeida, Janko Altenschmidt, Sam Altman, Shyamal Anadkat, et~al.
\newblock Gpt-4 technical report.
\newblock {\em arXiv preprint arXiv:2303.08774}, 2023.

\bibitem{anthropic2024claude3}
Anthropic.
\newblock The claude 3 model family: Opus, sonnet, haiku.
\newblock Model card, Anthropic, March 2024.

\bibitem{bordes2024introductionvisionlanguagemodeling}
Florian Bordes, Richard~Yuanzhe Pang, Anurag Ajay, Alexander~C. Li, Adrien Bardes, Suzanne Petryk, Oscar Mañas, Zhiqiu Lin, Anas Mahmoud, Bargav Jayaraman, Mark Ibrahim, Melissa Hall, Yunyang Xiong, Jonathan Lebensold, Candace Ross, Srihari Jayakumar, Chuan Guo, Diane Bouchacourt, Haider Al-Tahan, Karthik Padthe, Vasu Sharma, Hu~Xu, Xiaoqing~Ellen Tan, Megan Richards, Samuel Lavoie, Pietro Astolfi, Reyhane~Askari Hemmat, Jun Chen, Kushal Tirumala, Rim Assouel, Mazda Moayeri, Arjang Talattof, Kamalika Chaudhuri, Zechun Liu, Xilun Chen, Quentin Garrido, Karen Ullrich, Aishwarya Agrawal, Kate Saenko, Asli Celikyilmaz, and Vikas Chandra.
\newblock An introduction to vision-language modeling, 2024.

\bibitem{choi2016convolutionalrecurrentneuralnetworks}
Keunwoo Choi, George Fazekas, Mark Sandler, and Kyunghyun Cho.
\newblock Convolutional recurrent neural networks for music classification, 2016.

\bibitem{deshmukh2024pengiaudiolanguagemodel}
Soham Deshmukh, Benjamin Elizalde, Rita Singh, and Huaming Wang.
\newblock Pengi: An audio language model for audio tasks, 2024.

\bibitem{dong2024surveyincontextlearning}
Qingxiu Dong, Lei Li, Damai Dai, Ce~Zheng, Jingyuan Ma, Rui Li, Heming Xia, Jingjing Xu, Zhiyong Wu, Baobao Chang, Xu~Sun, Lei Li, and Zhifang Sui.
\newblock A survey on in-context learning, 2024.

\bibitem{Drossos2019ClothoAA}
Konstantinos Drossos, Samuel Lipping, and Tuomas Virtanen.
\newblock Clotho: an audio captioning dataset.
\newblock {\em ICASSP 2020 - 2020 IEEE International Conference on Acoustics, Speech and Signal Processing (ICASSP)}, pages 736--740, 2019.

\bibitem{Du2022ASO}
Yifan Du, Zikang Liu, Junyi Li, and Wayne~Xin Zhao.
\newblock A survey of vision-language pre-trained models.
\newblock In {\em International Joint Conference on Artificial Intelligence}, 2022.

\bibitem{dubey2024llama}
Abhimanyu Dubey, Abhinav Jauhri, Abhinav Pandey, Abhishek Kadian, Ahmad Al-Dahle, Aiesha Letman, Akhil Mathur, Alan Schelten, Amy Yang, Angela Fan, et~al.
\newblock The llama 3 herd of models.
\newblock {\em arXiv preprint arXiv:2407.21783}, 2024.

\bibitem{ghosh2024gamalargeaudiolanguagemodel}
Sreyan Ghosh, Sonal Kumar, Ashish Seth, Chandra Kiran~Reddy Evuru, Utkarsh Tyagi, S~Sakshi, Oriol Nieto, Ramani Duraiswami, and Dinesh Manocha.
\newblock Gama: A large audio-language model with advanced audio understanding and complex reasoning abilities, 2024.

\bibitem{gong2021astaudiospectrogramtransformer}
Yuan Gong, Yu-An Chung, and James Glass.
\newblock Ast: Audio spectrogram transformer, 2021.

\bibitem{hershey2017cnnarchitectureslargescaleaudio}
Shawn Hershey, Sourish Chaudhuri, Daniel P.~W. Ellis, Jort~F. Gemmeke, Aren Jansen, R.~Channing Moore, Manoj Plakal, Devin Platt, Rif~A. Saurous, Bryan Seybold, Malcolm Slaney, Ron~J. Weiss, and Kevin Wilson.
\newblock Cnn architectures for large-scale audio classification, 2017.

\bibitem{jiang2024effectivenessassessmentrecentlarge}
Yao Jiang, Xinyu Yan, Ge-Peng Ji, Keren Fu, Meijun Sun, Huan Xiong, Deng-Ping Fan, and Fahad~Shahbaz Khan.
\newblock Effectiveness assessment of recent large vision-language models, 2024.

\bibitem{piczak2015esc}
Karol~J Piczak.
\newblock Esc: Dataset for environmental sound classification.
\newblock In {\em Proceedings of the 23rd ACM international conference on Multimedia}, pages 1015--1018, 2015.

\bibitem{Pons2017EndtoendLF}
Jordi Pons, Oriol Nieto, Matthew Prockup, Erik~M. Schmidt, Andreas~F. Ehmann, and Xavier Serra.
\newblock End-to-end learning for music audio tagging at scale.
\newblock {\em ArXiv}, abs/1711.02520, 2017.

\bibitem{tang2024salmonngenerichearingabilities}
Changli Tang, Wenyi Yu, Guangzhi Sun, Xianzhao Chen, Tian Tan, Wei Li, Lu~Lu, Zejun Ma, and Chao Zhang.
\newblock Salmonn: Towards generic hearing abilities for large language models, 2024.

\bibitem{gemini}
Gemini Team, Rohan Anil, Sebastian Borgeaud, Jean-Baptiste Alayrac, Jiahui Yu, Radu Soricut, Johan Schalkwyk, Andrew~M Dai, Anja Hauth, Katie Millican, et~al.
\newblock Gemini: a family of highly capable multimodal models.
\newblock {\em arXiv preprint arXiv:2312.11805}, 2023.

\bibitem{yue2024mmmumassivemultidisciplinemultimodal}
Xiang Yue, Yuansheng Ni, Kai Zhang, Tianyu Zheng, Ruoqi Liu, Ge~Zhang, Samuel Stevens, Dongfu Jiang, Weiming Ren, Yuxuan Sun, Cong Wei, Botao Yu, Ruibin Yuan, Renliang Sun, Ming Yin, Boyuan Zheng, Zhenzhu Yang, Yibo Liu, Wenhao Huang, Huan Sun, Yu~Su, and Wenhu Chen.
\newblock Mmmu: A massive multi-discipline multimodal understanding and reasoning benchmark for expert agi, 2024.

\bibitem{zhang2024vision}
Jingyi Zhang, Jiaxing Huang, Sheng Jin, and Shijian Lu.
\newblock Vision-language models for vision tasks: A survey.
\newblock {\em IEEE Transactions on Pattern Analysis and Machine Intelligence}, 2024.

\end{thebibliography}

\newpage

\appendix
\section*{Appendix A: Spectrogram hyperparameters ablation study}

% \chris{an itemized list might make this easier to parse, space permitting}

% We varied key parameters such as amplitude scale (logarithmic or linear), colorscale (viridis or magma),  frequency scale (logarithmic or linear), spectrogram style transformations (raw amplitude magnitude spectrogram, mel spectrogram or mel-frequency cepstral coefficients (MFCCs)) and adding or removing the colorbar and labels to generate diverse spectrograms from the same audio files. We also tried low resolution spectrogram images as inputs by changing the `detail' parameter of GPT 4o so the images consume fewer tokens. The zero-shot and few-shot experiments cost about 40 cents and 4 dollars each using GPT-4o. The effect of all these hyperparameters is shown in \ref{fig:hyp}
We varied key parameters to generate diverse spectrograms from the same audio files:
\begin{itemize}
\item Amplitude scale: logarithmic or linear
\item Colorscale: viridis or magma
\item Frequency scale: logarithmic or linear
\item Spectrogram style: raw amplitude magnitude, mel spectrogram, or mel-frequency cepstral coefficients (MFCCs)
\item Visual elements: presence or absence of colorbar and labels
\item Image resolution: standard or low (adjusted via GPT-4o's 'detail' parameter)
\end{itemize}

Some variations (e.g., amplitude and frequency scales) are common in spectrogram-based classifiers, while others (e.g., colorscale, text labels) are more specific to VLMs. The aim was to assess how these variations in the spectrogram representation affected the model's ability to classify audio signals. Table \ref{table:experiment_results} presents the results of these experiments, showing the zero-shot VSC performance of GPT-4o on the ESC-10 dataset (first-fold) with different spectrogram hyperparameters. Notably, amplitude spectrograms with a linear frequency axis yielded the highest classification accuracy (35\%), suggesting that these settings better capture relevant audio features. The zero-shot performance on tuned parameters of all models is shown in the second column of the Table \ref{table:full}.

\begin{table}[h]
\caption{Zero-shot VSC performance of GPT-4o on ESC-10 (first fold) with various hyperparameters}

\label{table:experiment_results}
\centering
\begin{tabular}{lc}
\toprule
Parameter & Accuracy (\%) \\
\midrule
Default parameters & 27.50 \\
Linear frequency axis & 35.00 \\
Linear amplitude scale & 30.00 \\
Remove labels & 26.25 \\
Show colorbar & 23.75 \\
Magma colormap & 25.00 \\
Mel spectrogram & 25.00 \\
MFCCs & 13.75 \\
% 1s chunks & 22.50 \\
Low resolution & 20.00 \\
\bottomrule
\end{tabular}
\end{table}

\begin{figure}[h!]
    \centering
    \includegraphics[width=1\linewidth]{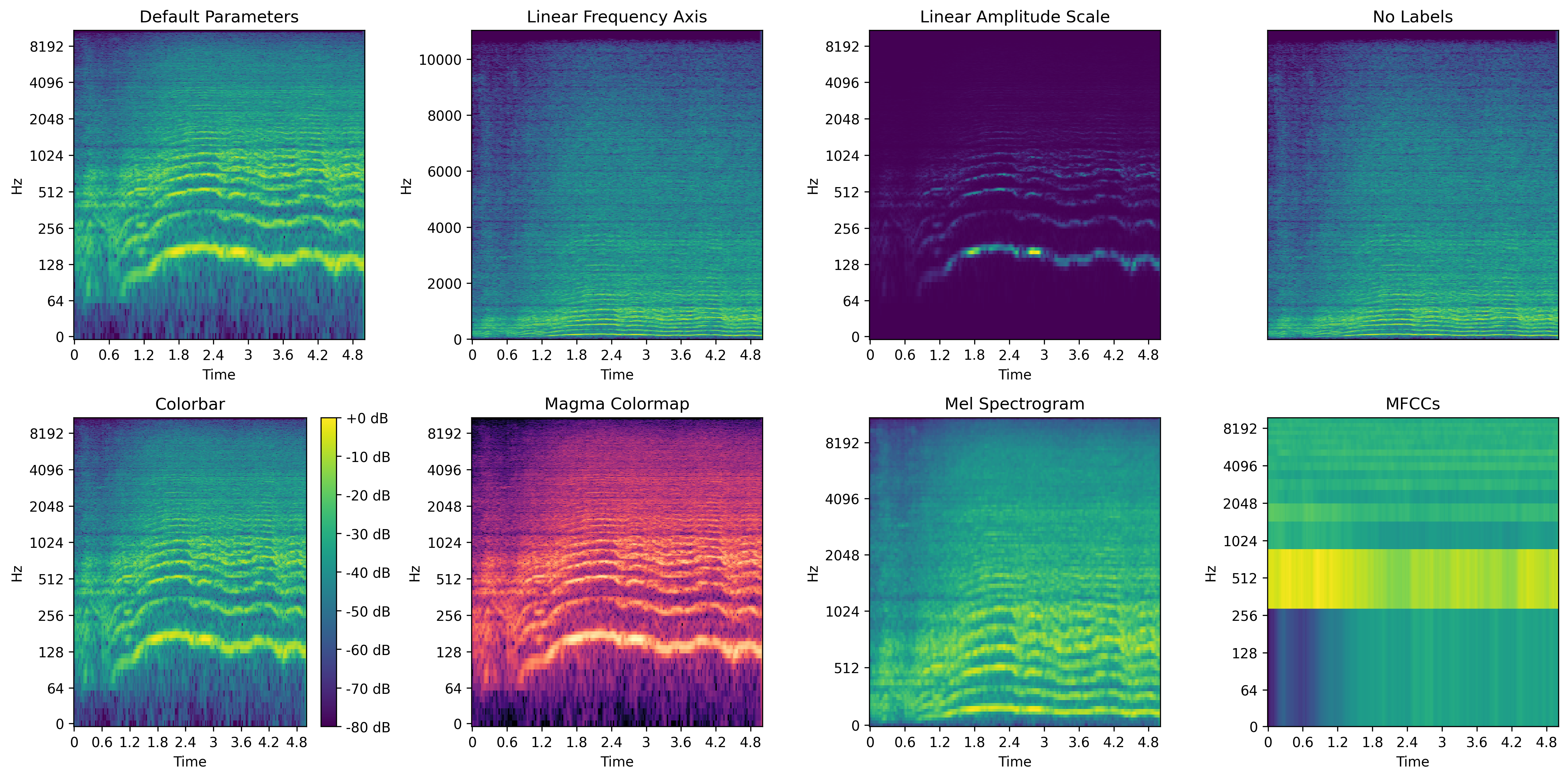}
    \caption{Example spectrograms for the same audio using different configurations described in the hyperparameter ablations section}
    \label{fig:hyp}
\end{figure}

% \section{Prompts for zero-shot and few-shot VSC}

% \documentclass{article}
% \usepackage{graphicx}
% \usepackage{caption}

% \section*{Appendix: System Prompt for Spectrogram Classification}
% \subsection{Prompts}
% We provide a prompt templates that we used for the experiments.

\section*{Appendix B: Prompts for VSC in zero-shot and few-shot settings}

Figures \ref{fig:zero-shot-prompt-template} and \ref{fig:few-shot-prompt-template} show the zero-shot and few-shot prompt templates for the visual spectrogram classification task.

\lstset{ 
    language=Python,            % choose the language of the code
    basicstyle=\ttfamily,        % the size of the fonts for the code
    keywordstyle=\color{blue},   % keyword color
    commentstyle=\color{orange},   % comment color
    stringstyle=\color{red},     % string color
    numbers=left,                % display line numbers
    numberstyle=\tiny,           % size of the numbers
    stepnumber=1,                % the step between two line numbers
    breaklines=true,             % automatic line breaking
    % frame=single                 % adds a frame around the code
}

\begin{figure*}[h!]
    \centering
    \begin{lstlisting}
ZERO_SHOT_PROMPT = """
{
    "role": "system",
    "content": "You are a helpful assistant with expertise in recognizing patterns and identifying classes based on visual representations of audio data."
},
{
    "role": "user",
    "content": [
        {
            "type": "text",
            "text": "Your task is to analyze a spectrogram, which is a visual representation of the frequency spectrum of sound over time, and determine the most likely sound class from a given list of possibilities. Analyze the spectrogram image, considering factors such as frequency patterns, intensity, and time variations. Focus solely on the patterns presented in the spectrogram. Do not let any assumptions about common sounds or environmental settings influence your decision. Here are the classes: ['dog', 'chainsaw', 'crackling_fire', 'helicopter', 'rain', 'crying_baby', 'clock_tick', 'sneezing', 'rooster', 'sea_waves']. Your response must always contain the exact name of the class only. For example, if you believe the spectrogram matches best with rain, your response would be rain. Here is the spectrogram:"
        },
        {
            "type": "image_url",
            "image_url": {
                "url": "data:image/png;base64,{image}"
            }
        }
    ]
}
    """
    \end{lstlisting}
    \caption{The template for prompting the VLM for zero-shot VSC. \{image\} is replaced by the spectrogram image to be classified.}
    \label{fig:zero-shot-prompt-template}
\end{figure*}

\begin{figure*}
    \centering
    \begin{lstlisting}
FEW_SHOT_PROMPT = """
{
    "role": "system",
    "content": "You are a helpful assistant with expertise in recognizing patterns and identifying classes based on visual representations of audio data."
},
{
    "role": "user",
    "content": [
        {
            "type": "text",
            "text": "Your task is to analyze spectrograms, which are visual representations of the frequency spectrum of sound over time, and determine the most likely sound class for a given spectrogram.\nHere are examples of spectrograms for different sound classes:"
        },
        {
            "type": "examples",
            "content": [
                {
                    "type": "text",
                    "text": "Spectrogram for {category-i}:"
                },
                {
                    "type": "image_url",
                    "image_url": {"url" "data:image/png;base64,{example-image-i}"}
                }
            ]
        },
        {
            "type": "text",
            "text": "\nNow, given a new spectrogram, analyze it considering factors such as frequency patterns, intensity, and time variations. Focus solely on the patterns presented in the spectrogram. Do not let any assumptions about common sounds or environmental settings influence your decision.\nYour task is to determine which of the example classes the new spectrogram most closely resembles. Your response must contain only the exact name of the class.\nHere is the new spectrogram to classify:"
        },
        {
            "type": "image_url",
            "image_url": {
                "url": "data:image/png;base64,{image}"
            }
        }
    ]
}
    """
    \end{lstlisting}
    \caption{The template for prompting the VLM for few-shot VSC. \{category-i\} and \{example-image-i\} are replaced by the name and example spectrogram for the i-th class where i will take values from 1 to n for n-shot classification and \{image\} is replaced by the spectrogram image to be classified.}
    \label{fig:few-shot-prompt-template}
\end{figure*}

% \noindent Below is an example placeholder for the spectrogram image.

% \begin{figure}[h]
%     \centering
%     \includegraphics[width=1\textwidth]{spectrogram_placeholder.png} % Replace with actual image path
%     \caption{Example Spectrogram for Classification Task}
%     \label{fig:spectrogram}
% \end{figure}

% \section{Example audio spectrograms in the ESC-10 dataset}

% Visual spectrogram classification (VSC) task involves classifying audio content based
% on the spectrogram representations. The spectrogram images of one example from each class of the ESC-10 dataset is shown in figure \ref{fig:example_spectrograms}. These time-frequency representations capture essential acoustic information about the audio.

% \section{Comparison of VLM and human expert performance on the VSC task}

% \begin{figure}
%     \centering
%     \includegraphics[width=1\linewidth]{CM.png}
%     \caption{Confusion matrices for GPT-4o and human experts (ensembled)}
%     \label{fig:CM}
% \end{figure}

\end{document}